\documentclass[12pt]{article}

\usepackage{graphicx}

\def\la{\mathrel{\mathpalette\fun <}}
\def\ga{\mathrel{\mathpalette\fun >}}
\def\fun#1#2{\lower3.6pt\vbox{\baselineskip0pt\lineskip.9pt
\ialign{$\mathsurround=0pt#1\hfil##\hfil$\crcr#2\crcr\sim\crcr}}}

\begin{document}

\title{{\bf Screening of the Coulomb potential in superstrong
magnetic field: atomic levels and spontaneous production of
positrons}}
\author{M.I.~Vysotsky \\
\small{\em A.I.~Alikhanov Institute of Theoretical and Experimental Physics,  117218 Moscow, Russia} \\ 
\small{\em All-Russian Scientific Research Institute of Automatics, 101000 Moscow, Russia}\\
\small{\em Moscow Engineering and Physics Institute, 115409 Moscow, Russia}\\
\small{\em Novosibirsk State University, 630090 Novosibirsk, Russia}}

\date{}  
\maketitle

\begin{abstract}
The expanded variant of the lectures delivered at 
the 39th ITEP Winter School in 2011.
\end{abstract}

\section{Introduction}

Loop corrections modify the Coulomb potential: electron loop insertion
into the photon propagator leads to the Uehling--Serber correction to
the electric potential of point-like nuclei \cite{BLP}. Though it leads
to an important phenomenon contributing to the Lamb shift of the
energies of atomic electrons numerically the shift being of the order of
$\alpha^5 m_e$  is small (we are using Gauss system of units,
where $\alpha = e^2 = 1/137$ and in all formulas $\hbar = c = 1$ is
implied).

Analogous correction in the case of external magnetic
field qualitatively change the behavior of atomic energies:
in particular the energy of the ground level remains finite in the
limit $B\to\infty$; also spontaneous production of positrons becomes
possible only for nuclei with $Z\geq 52$. Without taking radiative
corrections into account in the limit of infinite magnetic field
energy of ground atomic level tends to minus infinity and point-like
nucleus
with any $Z$ becomes critical at large enough $B$. At magnetic
fields $B > (Ze^2)^2 m_e^2/e$ the characteristic size of the
electron wave function in the transverse to the magnetic field direction
$a_H = 1/\sqrt{eB}$ (the so-called Landau radius) becomes smaller than
Bohr radius $a_B = 1/(Ze^2 m_e)$ making the Coulomb problem
essentially one-dimensional. Singularity of the Coulomb potential
in $d=1$ is stronger than in $d=3$. In $d=1$ the energy of the
ground level is unbounded from below: the ``fall to the center''
phenomenon occurs. In the case of external magnetic field the
singularity is cured by the finite value of $a_H$: at $|z|\la a_H$
the Coulomb problem remains three dimensional. This is the reason
why  ground level goes down when $B$ grows. At superstrong magnetic
fields $B\ga 3\pi m_e^2/e^3$ radiative corrections screen the
Coulomb potential at short distances $|z|\la 1/m_e$ and the
freezing of a ground state energy occurs: it remains finite at
$B\to\infty$ (\cite{SU}, \cite{V}, \cite{MV}). For $Z\geq 52$ the
value of freezing energy is below $-m_e$, so the ground level
enters lower continuum when $B$ increases and spontaneous
production of $e^+ e^-$ pair from vacuum becomes energetically
possible and thus takes place. In this process electron occupies
ground level while positron is emitted to infinity. For $Z < 52$
freezing energy is above $-m_e$ and spontaneous positron production
does not occur \cite{GMV}.

There exists the direct correspondence between radiative corrections
to the Coulomb potential in $d=3$ case in strong magnetic
field $B > B_0 \equiv m_e^2/e$ and radiative corrections to the
Coulomb potential in $d=1$ QED. That is why we start our
presentation (in Section 2) from the analysis of the Coulomb potential
in $D=2$ QED of massive fermions. When these fermions are light,
$g^2 > m^2$, the exponential screening of the Coulomb potential
at short distances occurs. In the limit $m\to 0$
(massless $D=2$ QED, the so-called Schwinger model) this
exponential screening occurs at all distances because photon gets
mass, $m_\gamma = 2g$ \cite{S}. In Section 3 we analyze radiative
corrections to the Coulomb potential in $D=4$ QED in external
magnetic field. The role of the coupling constant $g^2$ here
plays the product $e^3 B$, and for $e^3 B > m_e^2$ the screening
of the Coulomb potential occurs as well. In Section 4 the
structure of atomic levels on which the lowest Landau level
(LLL) in the presence of atomic nucleus splits is determined.
In Section 5 the Dirac equation for hydrogenlike ion at superstrong
magnetic field will be derived and effect of screening will
be studied for $Z=1$.
In Section 6 the influence of the screening of the Coulomb
potential on the values of critical nuclei charges is discussed.
In Section 7 the obtained results are summarized.

Let us finish the Introduction discussing the numerical values of
magnetic fields we are dealing with in these lectures.
The magnetic field at which the Bohr radius of a hydrogen
atom becomes equal to Landau radius is
$B_a = e^3$ $m_e^2 \approx 2 \cdot 10^9$ gauss, which is much
larger than a magnetic field ever made artificially on Earth:
$B_{\rm lab} \approx 3 \cdot 10^7$ gauss.
An interest to the atomic spectrum in the magnetic fields $B > B_a$
was triggered by the experiments with semiconductors, where
electron-hole bound system called exciton is formed.
Both effective charge and mass of electrons in
semiconductors are much lower than in vacuum making $B_a$
in kilogauss scale reachable.

The so-called Schwinger magnetic field
$B_0 = m_e^2/e \approx 4.4 \cdot 10^{13}$ gauss and magnetic
field at which the screening of the Coulomb potential
occurs $B \approx 3\pi m_e^2 /e^3 \approx 6 \cdot 10^{16}$ gauss
should be compared with the magnetic fields at
pulsars $\sim 10^{13}$ gauss and magnetars $\sim 10^{15}$ gauss.
Although the application of the results obtained in the
condensed matter physics (say graphen, where the mass of
charge carrier can be arbitrary low while the value of
charge approach one) can not be excluded, our main interest
in the problem considered is purely theoretical.

\section{The Coulomb potential in $D=2$ QED of massive fermions}

Summing up diagrams shown in Fig. 1 we get the following
formula for the potential of point-like charge:
\begin{equation}
\Phi(k) \equiv A_0(k) = -\frac{4\pi g}{k^2 + \Pi(k^2)} \; , \;\;
\Pi_{\mu\nu} \equiv (g_{\mu\nu}-\frac{k_\mu k_\nu}{k^2})\Pi(k^2) \;\; ,
\label{1}
\end{equation}
where $\Pi(k^2)$ is the one-loop expression for the photon polarization
operator. It can be obtained from the textbook expression for the polarization
operator calculated with the help of dimensional regularization \cite{AB}
in the limit $D\to 2$:
\begin{equation}
\Pi(k^2) = 4g^2\left[\frac{1}{\sqrt{t(1+t)}}\ln(\sqrt{1+t} +\sqrt
t) -1\right] \equiv -4g^2 P(t) \;\; ,  \label{2}
\end{equation}
where $t\equiv -k^2/4m^2, \;\; [g] = $mass.

\bigskip

\begin{center}
\bigskip
\includegraphics[width=.8\textwidth]{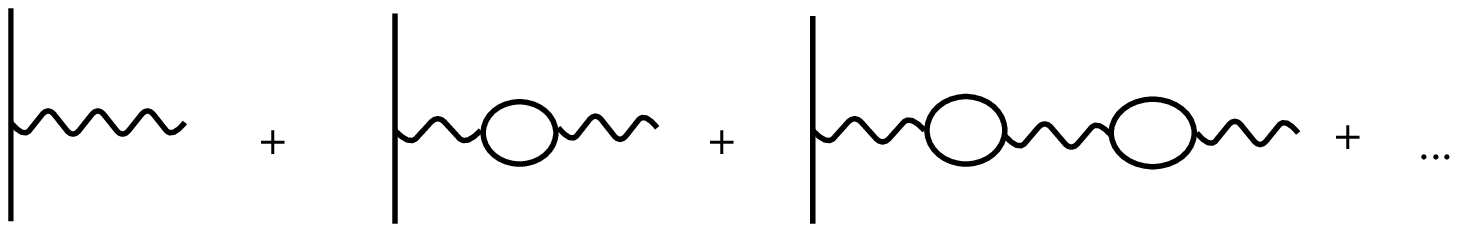}

Fig. 1. {\em Modification of the Coulomb potential due to the
dressing of the photon propagator.}

\end{center}

Let us note that $\Pi(k^2)$ is finite though corresponding integral
$\sim\int d^2 p/p^2$ is divergent in ultraviolet. The point is that
the trace of gamma matrices
which multiplies divergent integral is zero. In dimensional
regularization  the trace is proportional to $D-2$ while ultraviolet
divergency of integral over virtual momentum produces the factor
$1/(D-2)$ and the product of these two factors is finite.

In order to obtain an expression for the Coulomb potential in the
coordinate representation we take $k=(0, k_\parallel)$ and make the
Fourier transformation:
\begin{equation}
{\bf\Phi}(z) = 4\pi g \int\limits^\infty_{-\infty} \frac{e^{i
k_\parallel z} dk_\parallel/2\pi}{k_\parallel^2 + 4g^2
P(k_\parallel^2 /4m^2)} \;\; . \label{3}
\end{equation}
The potential energy for the charges $+g$ and $-g$ is
\begin{equation}
V(z) = -g\Phi(z) \;\; .
\label{4}
\end{equation}

The integral in (\ref{3}) cannot be expressed through elementary functions.
However it appears possible to find an interpolating formula for $P(t)$
which has good accuracy and is simple enough for the Fourier
transformation to be performed analytically.

The asymptotics of $P(t)$ are:
\begin{equation}
P(t) = \left\{
\begin{array}{lcl}
\frac{2}{3} t & , & t\ll 1 \\
1 & , & t\gg 1 \;\; .
\end{array}
\right.\label{5}
\end{equation}

Let us take as an interpolating formula for $P(t)$ the following
expression:
\begin{equation}
\overline{P}(t) = \frac{2t}{3+2t} \;\; .\label{6}
\end{equation}
The accuracy of this approximation is not worse than 10\% for the
whole interval of $t$ variation, $0 < t < \infty$. Substituting an
interpolating formula in (\ref{3}) we get:
\begin{eqnarray}
{\bf\Phi} & = & 4\pi g\int\limits^{\infty}_{-\infty} \frac{e^{i
k_\parallel z} d k_\parallel/2\pi}{k_\parallel^2 +
4g^2(k_\parallel^2/2m^2)/(3+k_\parallel^2/2m^2)} = \nonumber
\\
& = & \frac{4\pi g}{1+ 2g^2/3m^2}
\int\limits_{-\infty}^{\infty}\left[\frac{1}{k_\parallel^2} +
\frac{2g^2/3m^2}{k_\parallel^2 + 6m^2 + 4g^2}\right]
e^{ik_\parallel z} \frac{dk_\parallel}{2\pi} = \\
&=& \frac{4\pi g}{1+ 2g^2/3m^2}\left[-\frac{1}{2}|z| +
\frac{g^2/3m^2}{\sqrt{6m^2 + 4g^2}} {\rm exp}(-\sqrt{6m^2
+4g^2}|z|)\right] \;\; . \nonumber \label{7}
\end{eqnarray}

In the case of heavy fermions ($m\gg g$)  the potential is given
by the tree level expression; the corrections are suppressed as
$g^2/m^2$.

In the case of light fermions ($m \ll g$):

\begin{equation}
{\bf\Phi}(z)\left|
\begin{array}{l}
~~  \\
m \ll g
\end{array}
\right. = \left\{
\begin{array}{lcl}
\pi e^{-2g|z|} & , & z \ll \frac{1}{g} \ln\left(\frac{g}{m}\right) \\
-2\pi g\left(\frac{3m^2}{2g^2}\right)|z| & , & z \gg \frac{1}{g}
\ln\left(\frac{g}{m}\right) \;\; .
\end{array}
\right. \label{8}
\end{equation}
$m=0$ corresponds to Schwinger model; photon gets mass.

Light fermions make transition from $m>g$ to $m=0$ continuous.
In the case of light fermions the Coulomb potential in $D=2$ QED
is screened at distances $|z| \ga 1/(2g)$.

In Fig. 2 the potential energy for  $g=0.5,\;\; m=0.1$ is shown.
It is normalized to $V(0)=0$.
\bigskip

\begin{center}

\includegraphics[width=.8\textwidth]{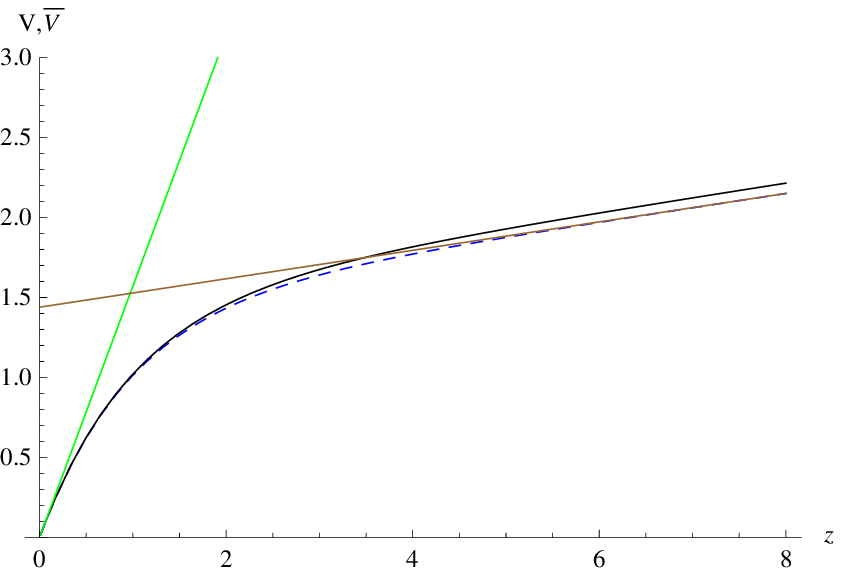}

Fig.2. { \em Potential energy of the charges $+g$ and $-g$ in $D=2$. The
solid curve corresponds to $P$; the dashed curve corresponds to
$\bar P$.}

\end{center}

\section{The Coulomb potential in $D=4$ QED in superstrong magnetic field}

In order to find the potential of a point-like charge we need an expression for photon polarization operator in the external magnetic field $B$. Long ago an expression for the electron propagator in constant and homogeneous external magnetic field was found by Schwinger \cite{Sch} as a parametric integral. For $B\gg B_0 \equiv m_e^2/e$ integration can be easily performed and compact expression for $G(k)$ follows. Using it one obtains an analytic expression for a photon polarization operator (see for example \cite{CR}).

To understand the reason for great simplification of the expression for the electron propagator in the limit $B\gg B_0$ one should start from the spectral representation of the propagator.
The solutions of Dirac equation in the homogeneous constant in time $B$ are known, so one can write the spectral representation of the electron Green function. The
denominators contain $k^2-m^2-2neB$, and for $B>>m^2/e$ and
$k_\parallel^2<<eB$ in sum over levels the lowest Landau level
(LLL, $n=0$) dominates. In the coordinate representation the
transverse part of LLL wave function is: $\Psi\sim
exp((-x^2-y^2)eB)$ which in the momentum representation gives
$\Psi\sim exp((-k_x^2-k_y^2)/eB)$ (we suppose that $B$ is directed
along the $z$ axis).

Substituting the electron Green functions we get the expression
for the polarization operator in superstrong $B$.

For $B>>B_0$, $k_\parallel^2 << eB$ the following expression is
valid \cite{SLS}:

\begin{eqnarray}
\Pi_{\mu\nu} & \sim & e^2 eB
\int\frac{dq_xdq_y}{eB}\exp(-\frac{q_x^2+q_y^2}{eB})* \nonumber \\
& * & \exp(-\frac{(q+k)_x^2+(q+k)_y^2}{eB})dq_0dq_z\gamma_{\mu}
\frac{1}{\hat q_{0,z}-m}(1-i\gamma_1 \gamma_2)\gamma_{\nu}*
\\
& * & \frac{1}{\hat q_{0,z}+\hat k_{0,z}-m}(1-i\gamma_1 \gamma_2)
= e^3 B * \exp(-\frac{k^2_\bot}{2eB}) *
\Pi_{\mu\nu}^{(2)}(k_\parallel\equiv k_z)\;\; . \nonumber
\label{9}
\end{eqnarray}

With the help of it the following result was obtained in \cite{MV}:

\begin{equation}
{\bf\Phi}(k) =\frac{4\pi e}{k_\parallel^2 + k_\bot^2 + \frac{2 e^3
B}{\pi} {\rm exp}\left(-\frac{k_\bot^2}{2eB}\right)
P\left(\frac{k_\parallel^2}{4m^2}\right)} \;\; , \label{10}
\end{equation}
\begin{eqnarray}
{\bf\Phi}(z) & = & 4\pi e \int\frac{e^{ik_\parallel z} d
k_\parallel d^2 k_\bot/(2\pi)^3}{k_\parallel^2 + k_\bot^2 +
\frac{2 e^3B}{\pi} {\rm
exp}(-k_\bot^2/(2eB))(k_\parallel^2/2m_e^2)/(3+k_\parallel^2/2m_e^2)}
= \nonumber \\
& = & \frac{e}{|z|}\left[ 1-e^{-\sqrt{6m_e^2}|z|} +
e^{-\sqrt{(2/\pi) e^3 B + 6m_e^2}|z|}\right] \;\; . \label{11}
\end{eqnarray}

For the magnetic fields $B \ll 3\pi m^2/e^3$ the potential is
Coulomb up to small power suppressed terms:
\begin{equation}
{\bf\Phi}(z)\left| \begin{array}{l}
~~  \\
e^3 B \ll m_e^2
\end{array}
\right. = \frac{e}{|z|}\left[ 1+ O\left(\frac{e^3
B}{m_e^2}\right)\right] \label{12}
\end{equation}
in full accordance with the $D=2$ case with the substitution $e^3B \rightarrow g^2$.

In the opposite case of the superstrong magnetic fields $B\gg 3\pi
m_e^2/e^3$ we get:
\begin{equation}
{\bf\Phi}(z) = \left\{
\begin{array}{lll}
\frac{e}{|z|} e^{(-\sqrt{(2/\pi) e^3 B}|z|)} \; , \;\;
\frac{1}{\sqrt{(2/\pi) e^3 B}}\ln\left(\sqrt{\frac{e^3 B}{3\pi
m_e^2}}\right)>|z|>\frac{1}{\sqrt{e B}}\\
\frac{e}{|z|}(1- e^{(-\sqrt{6m_e^2}|z|)}) \; , \;\;  \frac{1}{m_e} >
|z|
> \frac{1}{\sqrt{(2/\pi)e^3 B}}\ln\left(\sqrt{\frac{e^3 B}{3\pi
m_e^2}}\right) \\
\frac{e}{|z|}\;\; , \;\;\;\;\;\;\;\;\;\;\;\;\;\;\;\;\;\;\;\;\;\;\;
  |z| > \frac{1}{m_e}
\end{array}
\right. \;\; , \label{13}
\end{equation}

\begin{equation}
 \bar V(z) = - e{\bf\Phi}(z) \;\; .
\label{14}
\end{equation}

The Coulomb potential is screened at short distances $1/m_e > |z| > 1/\sqrt{e^3 B} \equiv a_H/e$.

In Fig. 3 the plot of a modified by the superstrong magnetic field Coulomb
potential as well as its short- and long-distance asymptotics are
presented.

\bigskip

\begin{center}

\includegraphics[width=.6\textwidth]{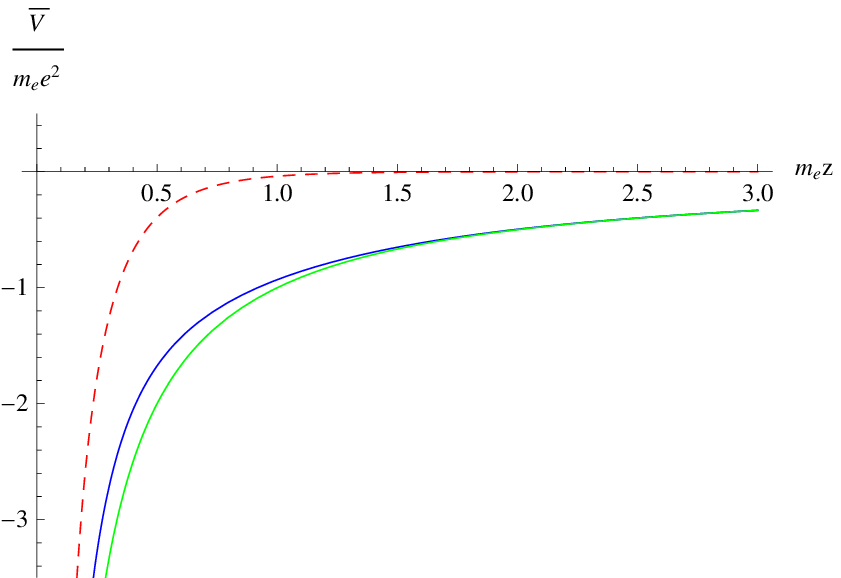}

Fig. 3. {\em The modified Coulomb potential at $B=10^{17}$G (blue,
dark solid) and its long distance (green, pale solid) and short
distance (red, dashed) asymptotics.}

\end{center}

Let us find the 3-dimensional shape of the screened Coulomb potential. The behavior of the potential in the transverse plane ($z=0$) can be found analytically in the limit $B\gg 3\pi m_e^2/e^3$ from the general expression:
\begin{equation}
\Phi(z,\rho) = 4\pi e\int\frac{e^{i\bar{k}_\bot \bar{\rho} +
ik_\parallel z} d k_\parallel d^2 k_\bot/(2\pi)^3}{k_\parallel^2 +
k_\bot^2 + \frac{2e^3 B}{\pi}
\exp(-\frac{k_\bot^2}{2eB})\frac{k_\parallel^2}{6m_e^2 +
k_\parallel^2}}  \label{141}
\end{equation}
neglecting exponent in the denominator, which is valid for $\rho\ga a_H$:
\begin{eqnarray}
\Phi(0,\rho)  = \left\{
\begin{array}{lll}
\frac{e}{\rho}\exp(-\sqrt{(2/\pi)e^3 B}\;\rho)& , & \rho <
\frac{1}{\sqrt{(2/\pi) e^3 B}}\ln\sqrt{\frac{e^3 B}{3\pi m_e^2}}
\\
\sqrt{\frac{3\pi m_e^2}{e^3 B}}\frac{e}{\rho}& , &
\frac{1}{\sqrt{(2/\pi) e^3 B}}\ln\sqrt{\frac{e^3 B}{3\pi m_e^2}} <
\rho \;\; ,
\end{array}
\right. \label{142}
\end{eqnarray}

and the Coulomb potential is screened at large $\rho$ in complete
analogy with the $D=2$ case.

For $|z| \gg 1/m_e$ in the integral (\ref{141}) the values
$|k_\parallel| \ll m_e$ dominate and we get:
\begin{equation}
\Phi(\rho, z)\left|_{z\gg 1/m_e}\right. = \frac{e}{\sqrt{z^2 +
(1+\frac{e^3 B}{3\pi m_e^2})\rho^2}} \;\; . \label{143}
\end{equation}

\bigskip

\begin{center}
\includegraphics[width=.4\textwidth]{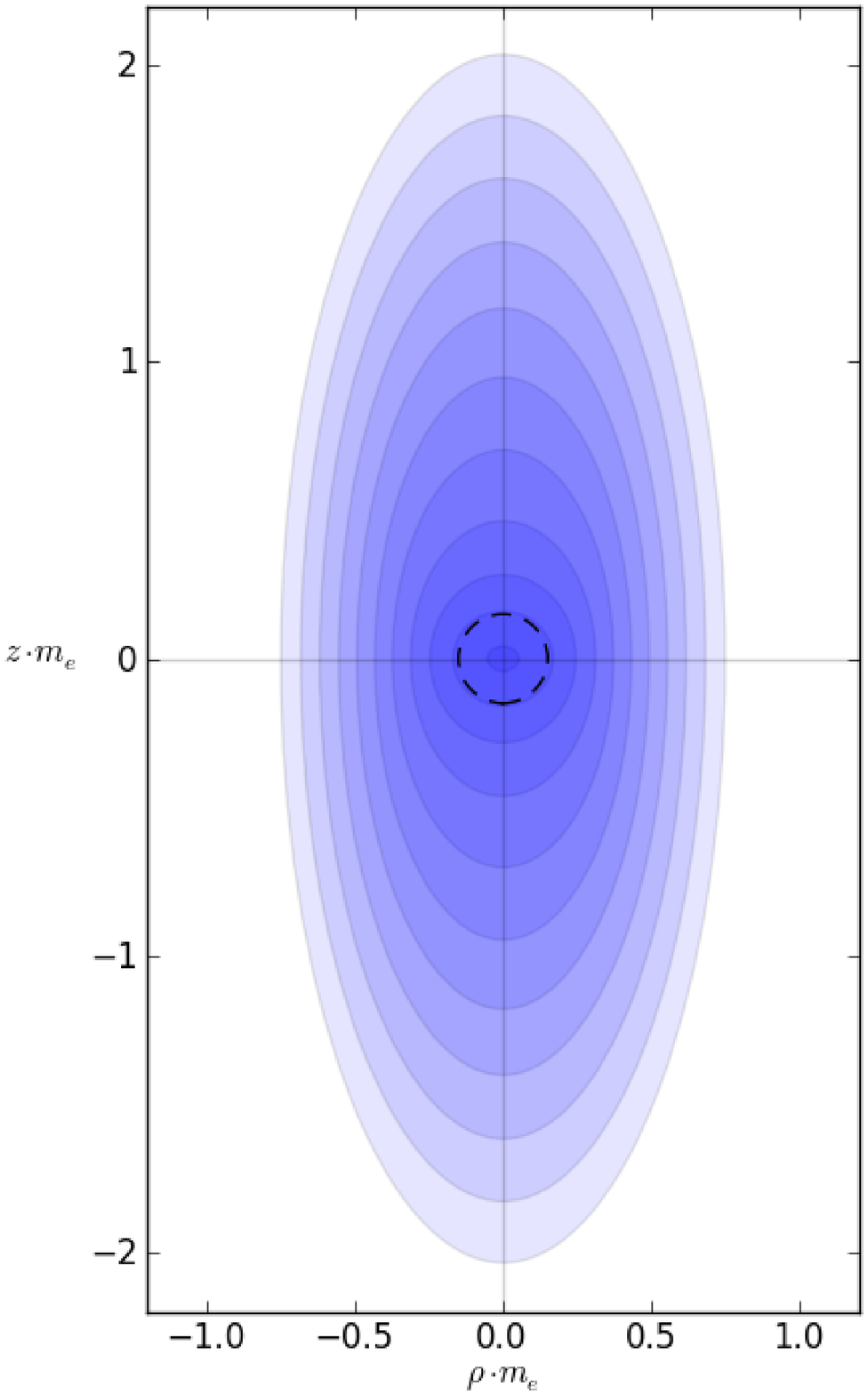}

\end{center}

\bigskip

Fig. 4. {\em The equipotential lines at $B/B_{0}=10^{4}$. The dashed line
    corresponds to $\sqrt{z^{2}+\rho^{2}}=\frac{1}{\sqrt{(2/\pi) e^3
        B}}\ln\sqrt{\frac{e^3 B}{3\pi m_e^2}}.$}

\bigskip

In Fig. 4 the equipotential lines are shown. The behavior of the screened
Coulomb potential in the transverse plane was found numerically in
\cite{SU}.

Finally for $3\pi m^2/e^3 \gg B \gg m^2/e$ expanding (\ref{143}) we
get:
\begin{equation}
\Phi(r) = \frac{e}{r}\left\{1-\frac{\alpha}{6\pi}(B/B_{\rm
0})\sin^2\theta\right\}\; , \;\; \theta = \bar r^\wedge\bar B \;
, \;\; r \equiv \sqrt{\rho^2 + z^2} \gg 1/m_{e} \;\; , \label{144}
\end{equation}
which coincides with the result obtained in \cite{SLS} where the
expression for the photon polarization operator at $B > B_0$ was
obtained as well.

The expression for the screened Coulomb potential was obtained
from the one-loop contribution to the photon
polarization operator in the external magnetic field $B\gg B_0$. In
momentum space it looks like:
\begin{equation}
\Phi(k_\parallel, k_0 = k_\bot = 0) = \frac{4\pi e}{k_\parallel^2
+ \frac{2e^3 B}{\pi}\frac{k_\parallel^2}{k_\parallel^2 + 6m_e^2}}
\;\; . \label{188}
\end{equation}

If higher loops contain the terms $\sim e^3 B(e^3
B/k_\parallel^2)^{n-1}$ they will drastically change the shape of
the potential in the coordinate space.

To calculate the radiative corrections one should use the electron
propagator in an external homogeneous magnetic field $B$. The
spectral representation of the electron propagator is a sum over
Landau levels and for $B \gg B_0$ the contribution of the lowest
level dominates \cite{SLS, Sm}:
\begin{equation}
G(k) = e^{-k_\bot^2/eB}(1-i\gamma_1 \gamma_2)\frac{\hat k_{0,3}
+ m_e}{k_{0,3}^2 - m_e^2} \;\; , \label{189}
\end{equation}
where the projector $(1-i \gamma_1 \gamma_2)$ selects the virtual electron
state with its spin oriented opposite to the direction of the magnetic
field $\bar B = (0,0,B)$. The contributions of the excited Landau levels
to $G$ yield a term in the denominator proportional to $eB$ and they
produce a correction of order $e^2 \equiv \alpha$ in the denominator
of (\ref{188}).

Two kinds of terms contribute to the polarization operator at the
two-loop level. First, there are terms in the electron propagators
which
represent the contributions of higher Landau levels. Just like in
the one-loop case they produce corrections suppressed as $e^2$ in the
denominator of (\ref{188}), i.e. terms of the order $e^5 B$
which can be safely neglected in comparison with the leading $\sim
e^3 B$ term. Second, there is the contribution from the leading term in
the electron propagator, given by (\ref{189}). Let us consider the
simplest diagram: the photon dressing of the electron propagator.
Neglecting the electron mass we get:
\begin{equation}
\gamma_\mu(1-i\gamma_1 \gamma_2)\hat k_{0,3}\gamma_\mu = -2[\hat
k_{0,3} - i\hat k_{0,3}\gamma_2\gamma_1]= -2\hat
k_{0,3}(1+i\gamma_1 \gamma_2) \;\; , \label{190}
\end{equation}
which gives zero when multiplying external propagator of electron,
since $(1+i\gamma_1 \gamma_2)(1-i\gamma_1 \gamma_2)=0$. This
result is a manifestation of the following well-known fact: in
massless QED in $D=2$ (Schwinger model) all loop diagrams are zero
except the one-loop term in the photon polarization operator
(see for example \cite{VB}). That is
why the contributions of the second kind are of the order of $\alpha
(e^{3}B)
\left(m^{2}_{e}k_{\parallel}^{2}/(k_{\parallel}^{2}+m_{e}^{2})^{2}\right)$ and
they are not important.

The generalization of the above arguments to higher loops is
straightforward. Let us note that absence of higher loop corrections
to polarization operator in Schwinger model is related to the absence
of renormalization of axial anomaly by higher loops. In $D=2$
anomaly is given by correlator of two currents and axial current is proportional to vector current (see for example \cite{PS}).

\section{Atomic levels in superstrong $B$}

We are interested in the spectrum of a hydrogen-like ion in a very strong magnetic field $B$. We will write all formulas for hydrogen since their generalization for $Z>1$ is straightforward.

In the absence of magnetic field
the spatial size of the wave function of the ground state atomic
electron is
characterized by the Bohr radius $a_B = 1/(m_e e^2)$, its energy
equals $E_0 = -m_e e^4/2 \equiv -Ry$, where $Ry$ is the Rydberg
constant. The transverse (with respect to $B$)
size of the ground state of the electron wave
function in an external magnetic field $B$ is characterized
by the Landau radius $a_H =
1/\sqrt{eB}$. The Larmour frequency of the electron precession is
$\omega_L = eB/m_e$. For a  magnetic field $B_a = e^3 m_e^2 = 2.35
\cdot 10^9$ gauss called ``atomic magnetic field'',
these sizes and energies are close to
each other: $a_B = a_H$, $E_0 \sim \omega_L$. We wish to study the
spectrum of the hydrogen atom in magnetic fields much larger than
$B_a$. In this case the motion of the electron is mainly controlled
by the
magnetic field: it makes many oscillations in this field before it
makes one in the Coulomb field of the nucleus. This is the condition for
applicability of the adiabatic approximation, used for this
problem for the first time in \cite{SS}.

The spectrum of a Dirac electron in a pure magnetic field is well
known \cite{BLP2}; it admits a continuum of energy levels due to the free
motion along the field:
\begin{equation}
\varepsilon_n^2 = m_e^2 + p_z^2 + (2n+1 + \sigma_z) eB  \;\; ,
\label{15}
\end{equation}
where $n = 0, 1,2, ...$; $\sigma_z = \pm 1$ is
the spin projection of the electron on $z$ axis
multiplied  by two. For magnetic fields
larger than $B_0 = m_e^2/e$, the electrons are relativistic with
only one exception: electrons belonging to the lowest Landau level (LLL,
$n=0$, $\sigma_z = -1$) can be non-relativistic.

In what follows we
will study the spectrum of electrons from LLL
in the  Coulomb field of the proton  modified by the
superstrong $B$. The solution can be found in \cite{LL}
of the
Schr\"{o}dinger equation for an electron in a constant in time homogeneous magnetic field $B$
in the gauge in which $\vec{A}  =
\frac{1}{2} \;\vec{B} \times \vec{r}$ in cylindrical coordinates ($\vec\rho, z$). The electron energies are:
\begin{equation}
E_{p_z n_\rho m \sigma_z} = \left(n_\rho + \frac{|m| +m+1+
\sigma_z}{2}\right)\frac{eB}{m_e} + \frac{p_z^2}{2m_e} \;\; ,
\label{16}
\end{equation}
where $n_\rho = 0,1,2,...$ is the number of nodal surfaces,
$m = 0, \pm 1, \pm 2, ...$ is the electron
orbital momentum projection on the  $z$ axis (direction of the magnetic
field)
and $\sigma_z = \pm 1$.  According to \cite{LL}, the LLL
wave functions are:
\begin{equation}
R_{0m}(\vec\rho) = \left[\pi(2a_H^2)^{1+|m|} (|m|!)\right]^{-1/2}
\rho^{|m|}e^{\left(im\varphi - \rho^2/(4a_H^2)\right)} \;\; , \label{17}
\end{equation}
$$
\rho = |\vec{\rho}|, \;\;\;
\int|R_{0m}(\vec\rho)|^2 d^2 \rho = 1 \; , \;\;  m=0,-1,-2,...
$$

We should now take into account the electric potential of the atomic
nucleus located at $\vec{\rho} = z = 0$. For $a_H \ll a_B$
the adiabatic
approximation can be used and the wave function
should be looked for in the following
form:
\begin{equation}
\Psi_{n 0 m (-1)} = R_{0m}(\vec{\rho})\chi_n(z) \;\; , \label{18}
\end{equation}
where $\chi_n(z)$ is the solution of a Schr\"{o}dinger equation
for an electron motion along the direction of the magnetic field:
\begin{equation}
\left[-\frac{1}{2m_e} \frac{d^2}{d z^2} + U_{eff}(z)\right]
\chi_n(z) = E_n \chi_n(z) \;\; . \label{19}
\end{equation}
Without screening the effective potential is given by the
following formula:
\begin{equation}
U_{eff} (z) = -e^2\int\frac{|R_{0m}(\vec{\rho})|^2}{\sqrt{\rho^2 +
z^2}}d^2 \rho \;\; , \label{20}
\end{equation}
which becomes the Coulomb potential for $|z| \gg a_H$
\begin{equation}
U_{eff}(z) \left|
\begin{array}{l}
~~  \\
z \gg a_H
\end{array}
\right. = - \frac{e^2}{|z|} \;\;  \label{21}
\end{equation}
and is regular at $z=0$

\begin{equation}
U_{eff}(0)
 \sim - \frac{e^2}{|a_H|} \;\; . \label{22}
\end{equation}

To take screening into account we must use (\ref{11}) to modify
(\ref{20}) (see below). Since $U_{eff}(z) = U_{eff}(-z)$, the wave
functions are odd or even under reflection $z\to -z$; the ground
states (for $m=0, -1, -2, ...$) are described by even wave
functions.

In Fig. 5 the different scales important in the consideration of the hydrogen
atom in strong magnetic field are shown.

\begin{center}
\bigskip
\includegraphics[width=.8\textwidth]{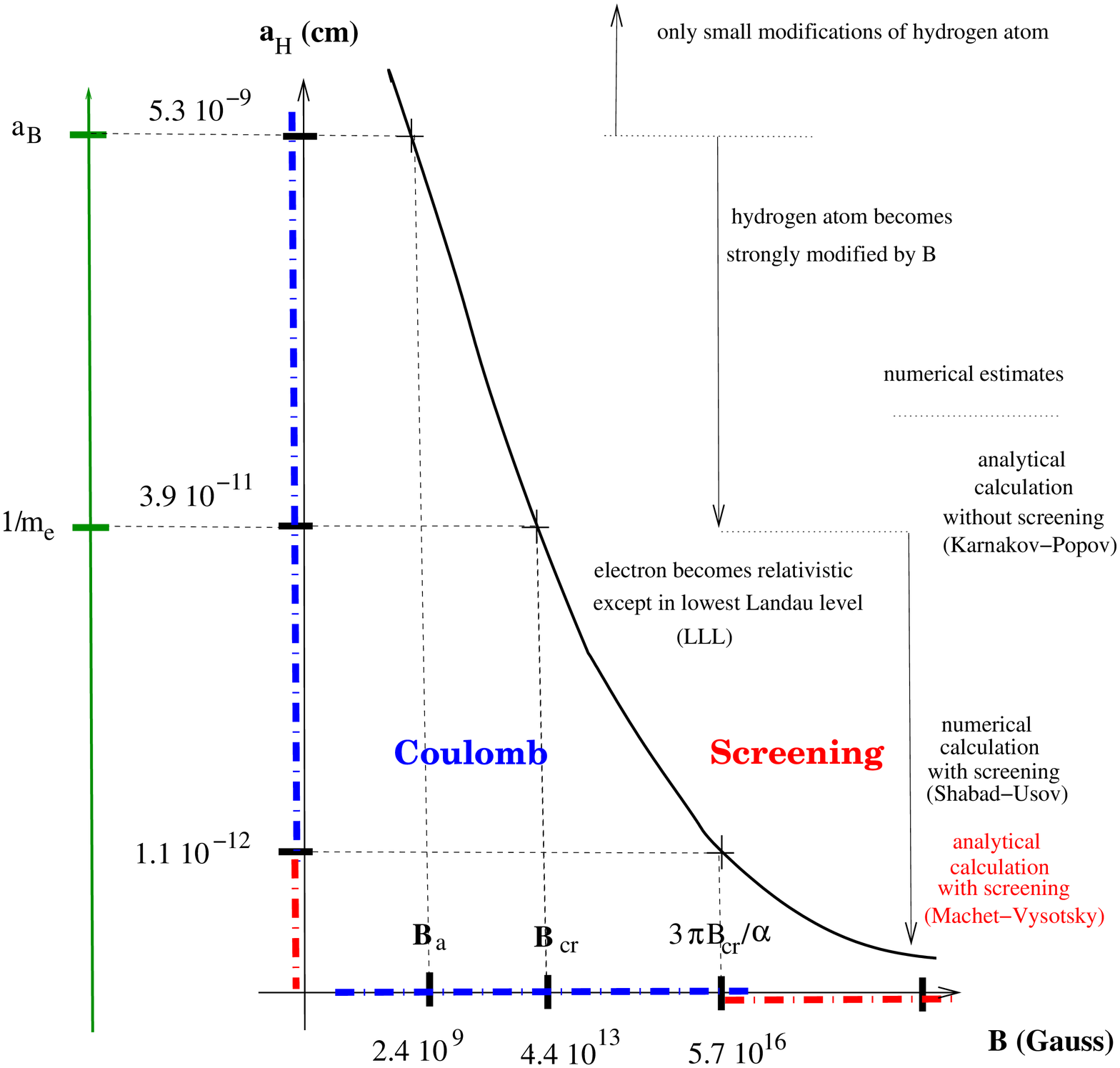}

Fig. 5. {\em Landau radius $a_H$ versus magnetic field $B$.}

\end{center}

To calculate the ground state of hydrogen atom in \cite{LL2}
the shallow-well
approximation is used:

\begin{equation}
E^{sw} = -2m_e\left[\int\limits_{a_H}^{a_B} U(z) dz\right]^2 =
-(m_e e^4/2)ln^2(B/(m_e^2e^3)) \label{27}
\end{equation}

Let us derive this formula. The starting point is the one-dimensional
Schr\"{o}dinger equation:
\begin{equation} -\frac{1}{2\mu}
\frac{d^2}{dz^2} \chi(z) + U(z) \chi(z) = E_0 \chi(z) \;\;
\label{28}
\end{equation}
Neglecting $E_0$ in comparison with $U$ and integrating (\ref{28}) we get:
\begin{equation}
\chi^\prime(a) = 2\mu\int\limits_0^a U(z)\chi(z) dz \;\; ,
\label{29}
\end{equation}
where we assume $U(z) = U(-z)$, that is why $\chi$ is even.

The next assumptions are: 1. the finite range of the potential
energy: $U(z) \neq 0$ for $a> z > -a$; 2. $\chi$ undergoes very
small variations inside the well. Since outside the well $\chi(z)
\sim e^{-\sqrt{2\mu |E_0|}\;z}$, we readily obtain:
\begin{equation}
|E_0| = 2\mu\left[\int\limits_0^a U(z) dz\right]^2 \;\; .
\label{30}
\end{equation}

For
\begin{equation}
\mu |U| a^2 \ll 1 \label{31}
\end{equation}
(condition for the potential to form a shallow well
which means that the absolute value of the energy of ground
level is much smaller than the absolute value of the potential
in the well) we get that,
indeed, $|E_0| \ll |U|$ and that the variation of $\chi$ inside
the well is small, $\Delta \chi/\chi \sim \mu |U|a^2 \ll 1$.

Concerning the one-dimensional Coulomb potential, it satisfies
this condition  only for $a\ll 1/(m_e e^2)\equiv a_B$.
This explains why the accuracy of $log^2$ formula (\ref{27}) is very
poor.

Much more accurate equation for atomic energies in strong magnetic
field was derived by B.M.Karnakov and V.S.Popov \cite{KP}.
It provides a
several percent accuracy for the energies of EVEN states for  $H >
10^3$ ($H \equiv  B/(m_e^2 e^3)$).

Main idea is to integrate Shr\"{o}dinger equation with effective
potential from $x=0$ till $x=z$, where $a_H<<z<<a_B$ and to equate
obtained expression for $\chi^\prime(z)/\chi(z)$ to the
logarithmic derivative of Whittaker function - the solution of
Shr\"{o}dinger equation with Coulomb potential, which
exponentially decreases at $z>>a_B$. In this way in \cite{KP}
the following equation was obtained:
\begin{eqnarray}
&&2\ln\left(\frac{z}{a_H}\right) + \ln 2 - \psi(1+|m|) + O(a_H/z)
= \nonumber \\
&&  2\ln\left(\frac{z}{a_B}\right) + \lambda + 2\ln \lambda +
2\psi\left(1-\frac{1}{\lambda}\right) + 4\gamma + 2\ln 2 +
O(z/a_B) \;\; , \label{32}
\end{eqnarray}

\begin{equation}
E = -(m_e e^4/2)\lambda^2 \;\; ,\label{33}
\end{equation}
where
$\psi(x)$ is the logarithmic derivative of the gamma function.

The energies of the ODD states are:
\begin{equation}
E_{\rm odd} = -\frac{m_e e^4}{2n^2} + O\left(\frac{m_e^2
e^3}{B}\right) \; , \;\; n = 1,2, ... \;\; . \label{34}
\end{equation}
So, for superstrong magnetic fields $B \sim m_e^2/e^3$ the
deviations of odd states energies
from the Balmer series are negligible.

From (\ref{32}) we get an equation for $\lambda$:
\begin{equation}
\ln(H) = \lambda + 2\ln\lambda +
2\psi\left(1-\frac{1}{\lambda}\right) + \ln 2 + 4\gamma +
\psi(1+|m|) \;\; , \label{36}
\end{equation}
where
$\psi(x)$ has simple poles at $x=0,-1,-2,...$.
So to reproduce large number at left hand side $\lambda$ should
be large (ground level) or follow Balmer series (excited levels).

When screening is taken into account an expression for effective
potential transforms into
\begin{equation}
\tilde U_{eff} (z) = -e^2\int  \frac{|R_{0m}(\vec{\rho})|^2}
{\sqrt{\rho^2 +z^2}} d^2\rho \left[1-e^{-\sqrt{6m_e^2}\;z} +
e^{-\sqrt{(2/\pi)e^3 B + 6m_e^2}\;z}\right]
 \;\;  \label{35}
\end{equation}

   Screening modifies the Coulomb potential at the distances
$|z|<1/m_e$ and since at these distances $m_e |U| a^2 = m_e e^2 a <
e^2<<1$, the approach leading to (\ref{32}) still works.

The modified Karnakov - Popov equation, which takes screening into account looks like:
\begin{equation}
\ln\left(\frac{H}{1+\displaystyle\frac{e^6}{3\pi}H}\right) =
\lambda + 2\ln\lambda + 2\psi\left(1-\frac{1}{\lambda}\right) +
\ln 2 + 4\gamma + \psi(1+|m|) \;\; . \label{37}
\end{equation}
 We see that at $H \equiv B/(m_e^2e^3
) \approx 3\pi /e^6 , B \approx 3\pi m_e^2/e^3$
freezing of the energies occur: left hand side of (\ref{37})
approach constant when $B$ further grows. In particular, for a ground state
at $B >> 3\pi m_e^2/e^3$ we obtain: $\lambda_0 = 11.2$, $E_0 = -1.7$ keV.

Energy levels on which LLL is splitted in the hydrogen atom
at $B >> 3\pi m_e^2/e^3$
are shown
on Fig. 6.
\bigskip

\begin{center}

\includegraphics[width=.48\textwidth]{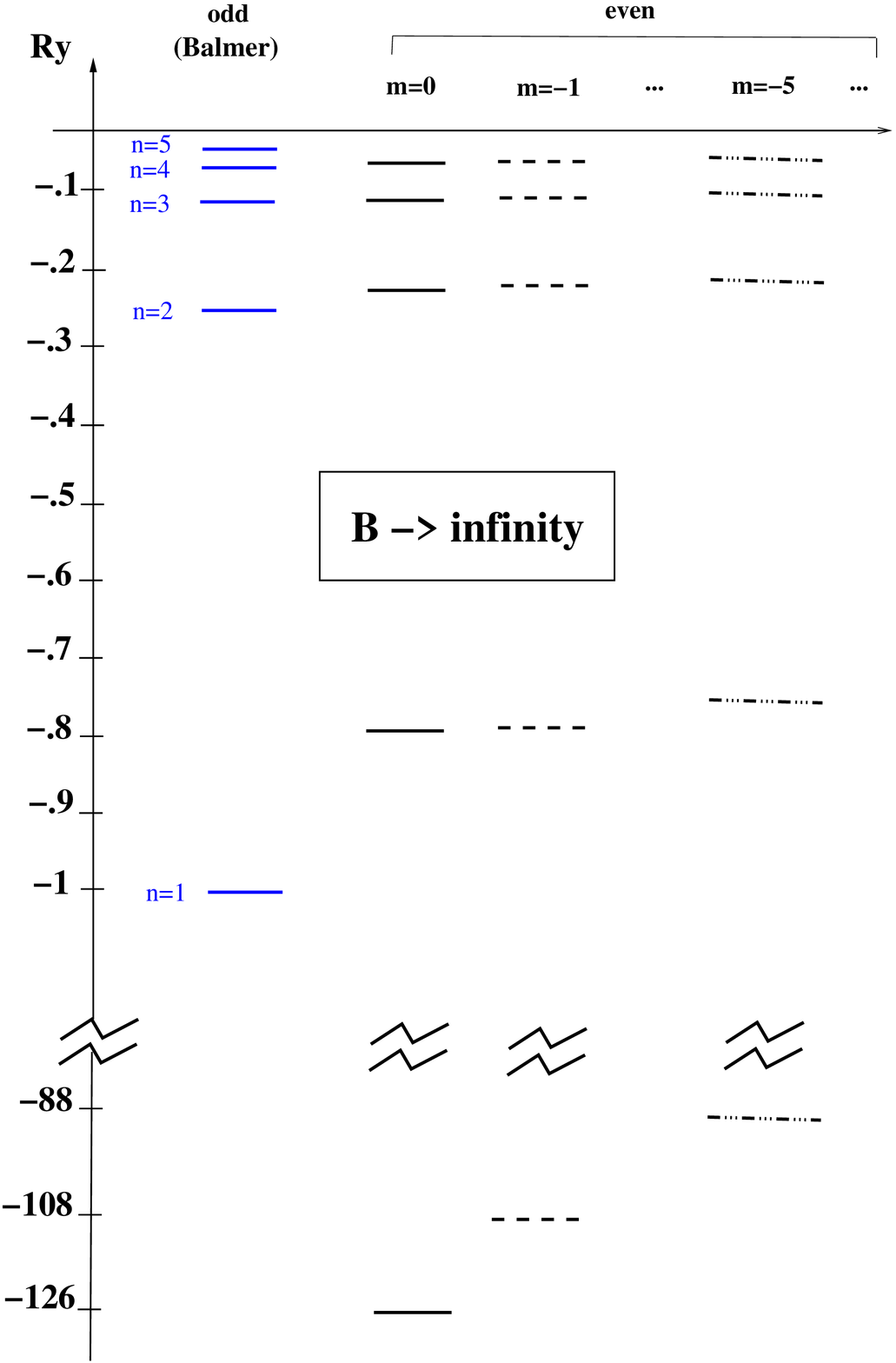}

\bigskip

Fig.6. {\em Spectrum of hydrogen levels in the limit
$B >> 3\pi m_e^2/e^3$. Energies are given in rydberg units, $Ry \equiv
13.6 \; eV$}.

\bigskip

\end{center}

\section{Dirac equation with a screened Coulomb potential, ${\bf Z=1}$}

In the previous Section the spectrum of energies on which the lowest Landau
level (LLL) splits in the proton electric field
was found by solving the corresponding Schr\"{o}dinger equation.
Since the ground state energy of hydrogen in the
limit of infinite $B$ equals $E_0 = -1.7$ keV,  the use of
the nonrelativistic Schr\"{o}dinger equation is at least
selfconsistent. However,  the size $a_{H}$ of the electron wave function for
$B > m_e^2/e^3$ in the direction transverse to the magnetic field
is much smaller than the electron Compton wavelength,
$a_H\equiv 1/\sqrt{eB} < e/m_e\ll 1/m_{e}$, which makes the nonrelativistic approach a bit
suspicious ($a_{H} = 1/m_e$ for $B = B_0$). That is why in
this Section we will study the ground
state energy of the electron
in a hydrogen-like ion in the presence of an external magnetic field
by analyzing the Dirac equation. Without taking screening into account
this problem was considered in paper \cite{ORS} (see also \cite{G},
were results obtained in \cite{ORS} were reproduced),
soon after it was found that a hydrogen-like ion becomes critical at
$Z \approx 170$: the electron ground level sinks into the lower continuum
($\varepsilon_{0}<-m_{e}$) and the vacuum becomes unstable by
spontaneous $e^{+}e^{-}$ pairs  production. These results were
obtained by solving the Dirac equation for an electron moving in the
field of a nucleus of finite  radius. That the phenomenon of
criticality can be studied only in the framework of the Dirac equation is an
additional motivation for us to go from Schr\"{o}dinger to Dirac.

From the numerical solution of the Dirac equation for the
ground electron level of a hydrogen atom in the Coulomb
potential we will find that the corrections to the nonrelativistic
results are small and that the estimate $\delta E \equiv |E_0^{\rm D} -
E_0^{\rm Sch}| \sim (E_0^{\rm Sch})^2/m_e$ works well.

Let us parametrize bispinor which describes electron wave function
in the following way:
\begin{equation}
\Psi = \left(
\begin{array}{c}
\varphi \\
\chi
\end{array}
\right) \; , \;\; \varphi = \left(
\begin{array}{c}
c_1 \\
c_2
\end{array}
\right) \; , \;\; \chi = \left(
\begin{array}{c}
b_1 \\
b_2
\end{array}
\right) \; . \label{101}
\end{equation}

Substituting $\Psi$ in the Dirac equation for the electron in an
external electromagnetic field we obtain:
\begin{equation}
\left\{
\begin{array}{ll}
(\varepsilon - m - e\varphi)\left(
\begin{array}{c}
c_1 \\
c_2
\end{array}
\right) + (-i\bar\sigma \frac{\partial}{\partial \bar r} + e\bar A
\bar\sigma) & \left( \begin{array}{c} b_1 \\
b_2 \end{array} \right) = 0 \\
-(i\bar\sigma \frac{\partial}{\partial\bar r} -e \bar A
\bar\sigma)  \left(
\begin{array}{c}
c_1 \\
c_2
\end{array}
\right) + (\varepsilon + m -e\varphi) & \left(\begin{array}{c} b_1
\\
b_2
\end{array}
\right) =0
\end{array}
\right. \label{102}
\end{equation}

Taking vector potential which describes constant magnetic field
$B$ directed along $z$ axis in the form $\bar A = (-\frac{1}{2} By ,
\frac{1}{2}Bx, 0)$, we get:
\begin{equation}
e\bar A \bar\sigma = -\frac{e}{2}B\left(
\begin{array}{cc}
0 & y + ix \\
y - ix & 0
\end{array}
\right) = -\frac{i}{2}eB\rho \left(\begin{array}{cc} 0 &
e^{-i\theta} \\
-e^{i\theta} & 0
\end{array}
\right) \;\; , \label{103}
\end{equation}
where $\rho = \sqrt{x^2 + y^2}$, $\theta \equiv \arctan (y/x)$.
Analogously we obtain:
\begin{equation}
-i\bar\sigma \frac{\partial}{\partial \bar r} = -i\left(
\begin{array}{cc}
\frac{\partial}{\partial z} &
e^{-i\theta}\frac{\partial}{\partial\rho}-\frac{ie^{-i\theta}}{\rho}
\frac{\partial}{\partial\theta} \\
e^{i\theta}\frac{\partial}{\partial\rho}+\frac{ie^{i\theta}}{\rho}
\frac{\partial}{\partial\theta} & -\frac{\partial}{\partial z}
\end{array}
\right) \;\; . \label{104}
\end{equation}
Substituting two last expressions in the Dirac equation we get:
\small
\begin{equation}
\left\{
\begin{array}{ll}
(\varepsilon - m - e\varphi)\left(\begin{array}{c} c_1 \\
c_2
\end{array}
\right) & -i\left(\begin{array}{cc} \frac{\partial}{\partial z} &
e^{-i\theta}\left(\frac{1}{2}eB\rho +
\frac{\partial}{\partial\rho}- \frac{i}{\rho}
\frac{\partial}{\partial\theta}\right) \\
e^{i\theta}\left(-\frac{1}{2}eB\rho +
\frac{\partial}{\partial\rho}+ \frac{i}{\rho}
\frac{\partial}{\partial\theta}\right) & -\frac{\partial}{\partial
z}
\end{array} \right) \left(
\begin{array}{c}
b_1 \\
b_2
\end{array}
\right) = 0 \\
(\varepsilon + m - e\varphi)\left(\begin{array}{c} b_1 \\
b_2
\end{array}
\right) & -i\left(\begin{array}{cc} \frac{\partial}{\partial z} &
e^{-i\theta}\left(\frac{1}{2}eB\rho +
\frac{\partial}{\partial\rho}- \frac{i}{\rho}
\frac{\partial}{\partial\theta}\right) \\
e^{i\theta}\left(-\frac{1}{2}eB\rho +
\frac{\partial}{\partial\rho}+ \frac{i}{\rho}
\frac{\partial}{\partial\theta}\right) & -\frac{\partial}{\partial
z}
\end{array} \right) \left(
\begin{array}{c}
c_1 \\
c_2
\end{array}
\right) = 0
\end{array}
\right. \label{105}
\end{equation}

\normalsize

Axial symmetry of electromagnetic field allows to determine
$\theta$ dependence of the functions $c_i$ and $b_i$:
\begin{equation}
\left(
\begin{array}{c}
c_1 \\
c_2
\end{array}
\right) = \left(
\begin{array}{cc}
c_1(\rho, z) & e^{i(M-1/2)\theta} \\
c_2(\rho, z) & e^{i(M+1/2)\theta}
\end{array}
\right) \; , \;\; \left(
\begin{array}{c}
b_1 \\
b_2
\end{array}
\right) = \left(
\begin{array}{cc}
b_1(\rho, z) & e^{i(M-1/2)\theta} \\
b_2(\rho, z) & e^{i(M+1/2)\theta}
\end{array}
\right) \; , \label{106}
\end{equation}
where $M = \pm 1/2, \pm 3/2$, ... is the projection of electron
angular momentum on $z$ axis. Substituting (\ref{106}) in
(\ref{105}) we get four linear equations for four unknown
functions $c_i$ and $b_i$ (here and below $c_1 \equiv c_1(\rho,
z)$, $b_1 \equiv b_1(\rho, z)$ ...):
\begin{eqnarray}
(\varepsilon - m - e\varphi)c_1 + i(-b_{1z} -b_{2\rho} -
\frac{M+1/2}{\rho} b_2 - \frac{eB\rho}{2} b_2) = 0 \nonumber \\
(\varepsilon - m - e\varphi)c_2 + i(-b_{1\rho} +
\frac{M-1/2}{\rho} b_1 + \frac{eB\rho}{2} b_1 + b_{2z}) = 0 \nonumber \\
(\varepsilon + m - e\varphi)b_1 + i(-c_{1z} -c_{2\rho} -
\frac{M+1/2}{\rho} c_2 - \frac{eB\rho}{2} c_2) = 0 \nonumber \\
(\varepsilon + m - e\varphi)b_2 + i(-c_{1\rho} +
\frac{M-1/2}{\rho} c_1 + \frac{eB\rho}{2} c_1 + c_{2z}) = 0 \;\; ,
\label{107}
\end{eqnarray}
where $b_{1z} \equiv \partial b_1/\partial z$, $b_{1\rho} \equiv
\partial b_1/\partial \rho$, ...  Ground energy state has $s_z =
-1/2$, $l_z =0$. Taking $M=-1/2$ we should look for solution of
(\ref{107}) with $c_1 = b_1 =0$:
\begin{equation}
\left\{
\begin{array}{l}
b_{2\rho} + \frac{eB\rho}{2}b_2 = 0 \\
\\
c_{2 \rho} + \frac{eB\rho}{2}c_2 = 0 \;\; ,
\end{array}\right.\label{108}
\end{equation}
\begin{equation}
\left\{
\begin{array}{l}
(\varepsilon - m - e\varphi)c_2 + ib_{2z} = 0 \\
(\varepsilon + m - e\varphi)b_2 + ic_{2z} = 0 \;\; .
\end{array}\right.
\label{109}
\end{equation}
The dependence on $\rho$ is determined by (\ref{108}):
\begin{equation}
\left\{
\begin{array}{l}
b_2(\rho, z) = e^{-eB\rho^2/4}(-i)f(z) \\
c_2(\rho, z) = e^{-eB\rho^2/4}g(z) \;\; .
\end{array}
\right. \label{110}
\end{equation}

Substituting the last expressions in (\ref{109}) and averaging
over fast motion in transverse to the magnetic field plane we
obtain two first order differential equations which describes
electron motion along magnetic field in an effective potential
$\bar V(z)$:
\begin{equation}
  \label{111}
  \begin{array}{c}
  g_{z}-(\varepsilon+m_{e}-\bar{V})f=0,\\
  f_{z}+(\varepsilon-m_{e}-\bar{V})g=0,
\end{array}
\end{equation}

\begin{equation}
\bar V(z) = -\frac{Ze^2}{a_H^2}\int\limits_0^\infty
\frac{\exp\left(-\frac{\rho^2}{2a_H^2}\right)}
{\sqrt{\rho^2 + z^2}}\rho d\rho \;\; . \label{112}
\end{equation}

 At large
distances $|z|\gg a_H$ the effective potential equals Coulomb, and the
solutions of the equations (\ref{111}) exponentially decreasing at
$|z|\to\infty$ are linear combinations of the Whittaker functions. At
short distances the equations (\ref{111}) can be easily integrated for
$|\bar V(z)| \gg |\varepsilon \pm m_e|$ (as far as $|\varepsilon|<m_e$
condition for $|\bar V(z)|$ will be for sure valid for  $|\bar V(z)|>2m_e$,
which is equivalent to
the following inequality: $z \ll Z e^2/(2m_e)$), where they looks like:
\begin{equation}
g_z + \bar V f = 0 \; , \;\; f_z - \bar V g = 0 \;\; .
\label{1000}
\end{equation}
The result of the
integration is:
\begin{eqnarray}
g(z) = B_1 \cos w(z) + B_2 \sin w(z) \;\; , \nonumber \\
f(z) = B_1 \sin w(z) - B_2 \cos w(z) \;\; \label{1001}
\end{eqnarray}
where
\begin{equation}
w(z) = \int\limits^{z}_{0}\bar V(z')dz' \label{1002}
\end{equation}
and $B_1$, $B_2$ are normalization constants.

The functions $g(z)$ and $f(z)$ have opposite parities; for the
ground state $g(z)$ should be even, so $B_2 = 0$, and matching
logarithmic derivatives at the point $z_0$ we obtain:
\begin{equation}
-\bar V(z_0) \tan w(z_0) = \frac{d}{dz}\ln g(z_0) \;\; , \label{1003}
\end{equation}
\begin{equation}
  \label{113}
Z e^2/(2m_e)\gg z_0 \gg a_H
\end{equation}
Substituting proper combination of the Whittaker functions 
for $g(z)$ we obtain  an algebraic equation for the ground state 
energy (it coincides with
Eq. (22) in \cite{ORS} in the limit $R/a_{H}\ll 1$, where $R$ is the nucleus
radius):
\begin{eqnarray}
 Ze^2\ln\left(2\frac{\sqrt{m_e^2-\varepsilon^2}}{\sqrt{eB}}\right) +
\arg\Gamma\left(-\frac{Ze^2 \varepsilon}{\sqrt{m_e^2 -
\varepsilon^2}} + iZe^2\right) +\nonumber\\
+\arctan\left(\sqrt{\frac{m_e+\varepsilon}{m_e-\varepsilon}}\right) -
 \arg \Gamma(1+2iZe^2) -
 \frac{Ze^2}{2}(\ln 2 + \gamma) =\frac{\pi}{2}+ n\pi \;\; , \label{114}
\end{eqnarray}
where $\gamma = 0.5772...$ is the Euler constant, and the argument of the
gamma function is given by
\begin{equation}
\arg\Gamma(x+iy) = -\gamma y + \sum_{k=1}^\infty\left(\frac{y}{k}
- \arctan\frac{y}{x+k-1}\right) \;\; . \label{115}
\end{equation}
For the ground level at $\varepsilon > 0$ one should take $n=0$, while
for $\varepsilon < 0$ it should be changed to $n=-1$.

According to (\ref{114}) when the magnetic field increases the
ground state energy goes down and reaches the lower continuum.

A matching point exists only if $B \gg 4m_e^2/(e(Ze^2)^2)$ (see
(\ref{113})) and (\ref{114}) is valid only for
these values of the magnetic field.

Thus, without taking screening into account, from (\ref{114}) we can
obtain the dependence of the ground state energy of a hydrogen atom
on the magnetic field for $B \gg 4m_e^2/e^5$.
Screening modifies the Coulomb potential at distances
smaller than the electron Compton wave length, and from the
condition $|\bar V(1/m_e)| \gg 2 m_e$ we get $Ze^2>>2$. It means that
at $B > 3\pi m^2_e/e^3$ the phenomena of
screening does not allow to find analytically the ground state
energy.
In order to find the
ground state energy at $B \la 4m_e^2/e^5$ and to take screening
into account  the equations (\ref{111}) were solved
numerically. This system  can be transformed into one second order
differential equation for $g(z)$. By substituting
$g(z)=\left(\varepsilon+m_{e}-\bar{V}\right)^{1/2}\chi(z)$
a Schr\"{o}dinger-like equation for the function $\chi(z)$ was obtained
in \cite{ORS}:\footnote{This trick was exploited by V.S. Popov for
the qualitative analysis of the phenomenon of critical charge \cite{P}.}
\begin{equation}
\frac{d^2\chi}{dz^2} + 2m_{e}(E-U)\chi = 0 \;\; , \label{116}
\end{equation}

\begin{eqnarray*}
E = \frac{\varepsilon^2 - m_e^2}{2m_e} \; , \;\;
U = \frac{\varepsilon}{m_e}\bar V- \frac{1}{2m_e}\bar V^2 +
\frac{\bar V''}{4m_e(\varepsilon + m _e-\bar V)} + \frac{3/8(\bar
V')^2}{m_e(\varepsilon +m_e - \bar V)^2} \; ,
\end{eqnarray*}
where $\varepsilon$ is the energy eigenvalue of the Dirac equation and
$\bar V(z)$ is given in (\ref{112}). An equation (\ref{116})
was integrated numerically. Let us note that, while for $z \gg 1/m_e$
the last three
terms in the expression for $U$ are much smaller than the first
one (the only one remaining in the nonrelativistic approximation), at
$z \la 1/m_e$ the relativistic terms dominate and are very big for
$B\gg B_0$ at $z \sim a_H$ which makes numerical calculations very
complicated.

\begin{table}[h!]
\caption{\label{tab:1}Values of $\lambda$ for $Z=1$ without
  screening obtained from the Schr\"{o}dinger and Dirac
    equations. They start to differ substantially at enormous values
    of the magnetic field.}
\begin{center}
\begin{tabular}{||c||c|c|c|c||}
\hline
$B/B_{0}$ & KP-equation      & Numerical results   & Eq. (\ref{114}) & Numerical results \\
${~}$    & (Schr\"{o}dinger) &  (Schr\"{o}dinger)   & (Dirac)           & (Dirac)          \\
\hline
$10^{0}$  & 5.737     & 5.735   & 5.735   & 5.734          \\
$10^{1}$  & 7.374     & 7.374   & 7.370   & 7.371          \\
$10^{2}$  & 9.141     & 9.141   & 9.136   & 9.135          \\
$10^{3}$  & 11.00     & 11.00   & 10.99   & 10.99          \\
$10^{4}$  & 12.93     & 12.93   & 12.91   & 12.91          \\
$10^{5}$  & 14.91     & 14.91   & 14.88   & 14.88          \\
$10^{6}$  & 16.93     & 16.93   & 16.89   & 16.89          \\
$10^{7}$  & 18.98     & 18.98   & 18.93   & 18.92          \\
$10^{8}$  & 21.06     & 21.05   & 20.98   & 20.98          \\
$10^{9}$  & 23.16     & 23.15   & 23.05   & 23.05          \\
$10^{10}$ & 25.27     & 25.27   & 25.14   & 25.13          \\
$10^{11}$ & 27.40     & 27.40   & 27.23   & ${~}$          \\
$10^{12}$ & 29.54     & 29.54   & 29.33   & ${~}$          \\
\dots    & \dots     & ${~}$   & \dots   & ${~}$          \\
$10^{15}$ & 36.03     & ${~}$   & 35.64   & ${~}$          \\
\dots    & \dots     & ${~}$   & \dots   & ${~}$          \\
$10^{20}$ & 46.99     & ${~}$   & 46.11   & ${~}$          \\
\dots    & \dots     & ${~}$   & \dots   & ${~}$          \\
$10^{25}$ & 58.07     & ${~}$   & 56.40   & ${~}$          \\
\dots    & \dots     & ${~}$   & \dots   & ${~}$          \\
$10^{30}$ & 69.22     & ${~}$   & 66.38   & ${~}$          \\
\dots    & \dots     & ${~}$   & \dots   & ${~}$          \\
$10^{35}$ & 80.43     & ${~}$   & 75.98   & ${~}$          \\
\dots    & \dots     & ${~}$   & \dots   & ${~}$          \\
$10^{40}$ & 91.67     & ${~}$   & 85.10   & ${~}$          \\
\dots    & \dots     & ${~}$   & \dots   & ${~}$          \\
$10^{45}$ & 102.95    & ${~}$   & 93.67   & ${~}$          \\
\dots    & \dots     & ${~}$   & \dots   & ${~}$          \\
$10^{50}$ & 114.25    & ${~}$   & 101.62   & ${~}$          \\
\dots    & \dots     & ${~}$   & \dots   & ${~}$          \\
$10^{55}$ & 125.57    & ${~}$   & 108.89   & ${~}$          \\
\hline
\end{tabular}
\end{center}
\end{table}

In Table \ref{tab:1} the results for the ground state energy of a hydrogen atom
without screening are presented. The values of the magnetic field in
units of $B_0$ are given in the first column, while in columns
2-5 the values of $\lambda$ are given. By definition
\footnote{Let us note that the
definition of $\lambda$ used in \cite{ORS} differs from our:
$\lambda^{\mbox{\cite{ORS}}} \equiv e^2 \lambda$.}
\begin{equation}
E= \frac{\varepsilon^2 - m_e^2}{2m_e} \equiv -\frac{m_e e^4}{2}\lambda^2 \;\; . \;\;
\label{12}
\end{equation}

From Table \ref{tab:1} we see that:
\begin{enumerate}
\item  the results of the numerical integrations of the
  Schr\"{o}dinger and Dirac equations coincide within four
  digits:
  \begin{itemize}
  \item with the analytical Karnakov--Popov formula for the ground
    state energy ($n_\rho = m = 0$) \cite{KP} in the case of the
    Schr\"{o}dinger equation;
  \item with formula (\ref{114}) for $Z = 1$ in the case of the Dirac
    equation;
  \end{itemize}
\item for the relativistic shift of energy the following estimate
  works:
\end{enumerate}
\begin{equation}
E_{\rm Dirac} - E_{\rm Schr} \sim E_{\rm Schr} \frac{E_{\rm
Schr}}{m_e} \;\; , \label{118}
\end{equation}
$$
\delta\lambda \sim e^4 \lambda^3/4 \;\; .
$$

To take screening into account, the following
formula for the effective potential should be used in (\ref{116}) instead
of (\ref{112}):
\begin{eqnarray}
\bar V(z) = -\frac{Ze^2}{a_H^2}\left[1-e^{-\sqrt{6m_e^2}|z|} +
e^{-\sqrt{(2/\pi)e^3 B + 6m_e^2}|z|}\right]\int\limits_0^\infty
\frac{e^{-\rho^2/2a_H^2}}{\sqrt{\rho^2 + z^2}}\rho d \rho \;\; ,\qquad
\label{119}
\end{eqnarray}
where $Z=1$ for hydrogen.

The freezing of the ground state energy is due to a weaker singularity
of the potential with screening (\ref{119}) at $z\rightarrow 0$ for
$B\rightarrow \infty$ than that of the potential without screening
(\ref{112}). While the non-screened potential behaves like $1/z$ at small
$z$, the screened potential is proportional to $\delta(z)$ because,
when $B\rightarrow \infty$, the width of the region where it behaves
like $1/z$ shrinks to zero \cite{SU}.

In Table \ref{tab:2} the results of the analytical
formula for $\lambda$ with the account of screening  for the Schr\"{o}dinger equation (\ref{37}) are compared with the
results of the numerical integration of the Dirac equation. We see that in
the case of screening the relativistic shift of energy is also
very small, and due to it the ground state energies become a little
bit higher, just like without taking screening into account.
The freezing of the ground state energy occurs at $B/B_0 = 10^3 \div
10^4$, when $B \approx 3\pi m_e^2/e^3$.

\begin{table}[h!]
\caption{\label{tab:2}Values of $\lambda$ for $Z=1$ with screening.}
\begin{center}
\begin{tabular}{||c||c|c|c||}
\hline
$B/B_{0}$   & Eq. (\ref{37})  & Numerical results   &  Numerical results \\
${~}$ &       (Schr\"{o}dinger)             &  (Schr\"{o}dinger)  &   (Dirac)          \\
\hline
$10^{0}$  & 5.7     & 5.7   & 5.7       \\
$10^{1}$  & 7.4     & 7.4   & 7.4       \\
$10^{2}$  & 9.1     & 9.1   & 9.1       \\
$10^{3}$  & 10.5    & 10.6  & 10.6      \\
$10^{4}$  & 11.1    & 11.2  & 11.2      \\
$10^{5}$  & 11.2    & 11.3  & 11.3      \\
$10^{6}$  & 11.2    & 11.4  & 11.3      \\
$10^{7}$  & 11.2    & 11.4  & 11.3      \\
$10^{8}$  & 11.2    & 11.4  & 11.3      \\
\hline
\end{tabular}
\end{center}
\end{table}

\section{Screening versus critical nucleus charge}
\begin{table}[h!]
\caption{\label{tab:3}Values of $\varepsilon_{0}/m_{e}$ for $Z=40$.}
\begin{center}
\begin{tabular}{||c||c|c|c||}
\hline
$B/B_{0}$& Eq. (\ref{114}) & Numerical results   &  Numerical results \\
${~}$   &       (Dirac)      &  (Dirac)            &  with screening (Dirac)\\
\hline
$10^{0}$        & 0.819     & 0.850   & 0.850       \\
$10^{1}$        & 0.653     & 0.667   & 0.667       \\
$10^{2}$        & 0.336     & 0.339   & 0.346       \\
$10^{3}$        &-0.158     &-0.159   &-0.0765      \\
$10^{4}$        &-0.758     &-0.759   &-0.376       \\
$2\cdot 10^{4}$ &-0.926     &-0.927   &-0.423       \\
\cline{2-3}
\dots          &\multicolumn{2}{c|}{at $B/B_{0}\approx 2.85\cdot 10^{4},~ \varepsilon_{0}=-m_{e}$} & \dots \\
\cline{2-3}
$10^{5}$        & ---       & ---     &-0.488       \\
$10^{6}$        & ---       & ---     &-0.524       \\
$10^{7}$        & ---       & ---     &-0.535       \\
$10^{8}$        & ---       & ---     &-0.538       \\
\hline
\end{tabular}
\end{center}
\end{table}

According to \cite{ORS} nuclei with $Z\geq 40$ become critical in
an external $B$ (for smaller $Z$ the values of $a_H$ at which the
criticality is reached become smaller than the nucleus radius, the Coulomb
potential diminishes and thus the ground level does not reach the lower
continuum). 
\begin{table}[h!]
\caption{\label{tab:freezing_energy}Values of freezing ground state energies for different $Z$
    from the Schr\"{o}dinger and the Dirac equations. In order to find
  the freezing energies we must take $B/B_{0}\gg 3\pi/e^{2}$. In numerical calculations we took
  $B/B_{0}=10^{8}$.}
\centering
\begin{tabular}{||c||c|c||}
  \hline
  $Z$ & 
  $\left(E_{0}^{fr}\right)^{numerical}_{Schr}$, keV &
  $\left(\varepsilon_{0}^{fr}-m_{e}\right)^{numerical}_{Dirac}$, keV \\
  \hline
    1    &     -1.7  & -1.7 \\
    10   &     -88   & -87  \\
    20   &     -288  & -273 \\
    30   &     -582  & -519 \\
    40   &     -966  & -787 \\    
    49   &     -     & -1003\\    
  \hline
\end{tabular}
\end{table}

In Table \ref{tab:3} one can
see the dependence of the ground state electron energy $\varepsilon_{0}$
on the external magnetic field for $Z=40$. The numerical solutions of
(\ref{116}) are in good correspondence with the values of
$\varepsilon_{0}$ obtained from (\ref{114}). The numerical results with
screening are shown in the last column; we see that freezing
occurs in the relativistic domain $\varepsilon_0 \approx -m_e/2$
and the ground level never reaches lower continuum, $\varepsilon_0
> -m_{e}$.

In Table \ref{tab:freezing_energy}  we compare freezing energies for
different $Z$ obtained numerically from the nonrelativistic
Schr\"{o}dinger equation and from the Dirac equation. We see that for
$Z > 20$ the freezing occurs in the relativistic regime, where the
Schr\"{o}dinger equation should not be used. Let us stress that
  the value of $B$ at which the freezing occurs does not depend on $Z$.

From (\ref{114}) we obtain in the limiting case $\varepsilon\rightarrow
-m_{e}$ an equation which defines the value of the magnetic field at
which a nucleus with charge $Z$ becomes critical without taking screening
into account (it coincides with Eq. (32) from \cite{ORS}):
\begin{equation}
  \label{eq:ORS32}
  \frac{B}{B_{0}}=2(Z_{cr}e^{2})^{2}\exp\left(-\gamma+\frac{\pi-2\arg\Gamma
        (1+2iZ_{cr}e^{2})}{Z_{cr}e^{2}}\right).
\end{equation}
This equation is used to calculate the numbers in the second column of
Table~\ref{table4}.

\begin{table}[h!]
\caption{\label{table4}Values of $B/B_{0}$ at which
  $\varepsilon_{0}=-m_{e}$ according to the Dirac equation and nuclei
  become supercritical without (column 2,3) and with (column 4)
  taking screening into account.}
\begin{center}
\begin{tabular}{||c||c|c|c||}
\hline
$Z_{cr}$   & Eq. (\ref{eq:ORS32}) & Numerical results   &  Numerical results \\
${~}$ &       ${~}$                            &  without screening  &  with screening\\
\hline
90    &  118   &  116   &  122    \\
85    &  157   &  154   &  164    \\
80    &  213   &  210   &  229    \\
75    &  301   &  297   &  335    \\
70    &  444   &  438   &  527    \\
65    &  689   &  681   &  923    \\
60    &  1144  &  1133  &  1964   \\
55    &  2068  &  2053  &  6830   \\
54    &  2357  &  2340  &  10172  \\
53    &  2699  &  2681  &  17012  \\
52    &  3107  &  3087  &  35135  \\
51    &  3594  &  3572  &  $1.20\cdot 10^{5}$  \\
50    &  4181  &  4157  &  $1.14\cdot 10^{7}$  \\
45    &  9826  &  9787  &  ---  \\
40    &  28478 &  28408 &  ---  \\
35    &$1.12\cdot 10^{5}$& $1.12\cdot 10^{5}$  & ---  \\
30    &$6.99\cdot 10^{5}$& $6.98\cdot 10^{5}$  & ---  \\
25    &$9.27\cdot 10^{6}$& $9.27\cdot 10^{6}$  & ---  \\
\hline
\end{tabular}
\end{center}
\end{table}

From Table \ref{table4} we see that with the account of screening only the atoms with $Z \ga 52$ become
supercritical at the values of $B/B_0$ shown in the fourth column.
Because of screening a larger $B$ is needed for a nucleus to become
supercritical and the nuclei with $Z < 52$ never reach
supercriticality. This phenomenon is illustrated in Fig. 7.

\bigskip

\begin{center}
\bigskip
\includegraphics[width=.8\textwidth]{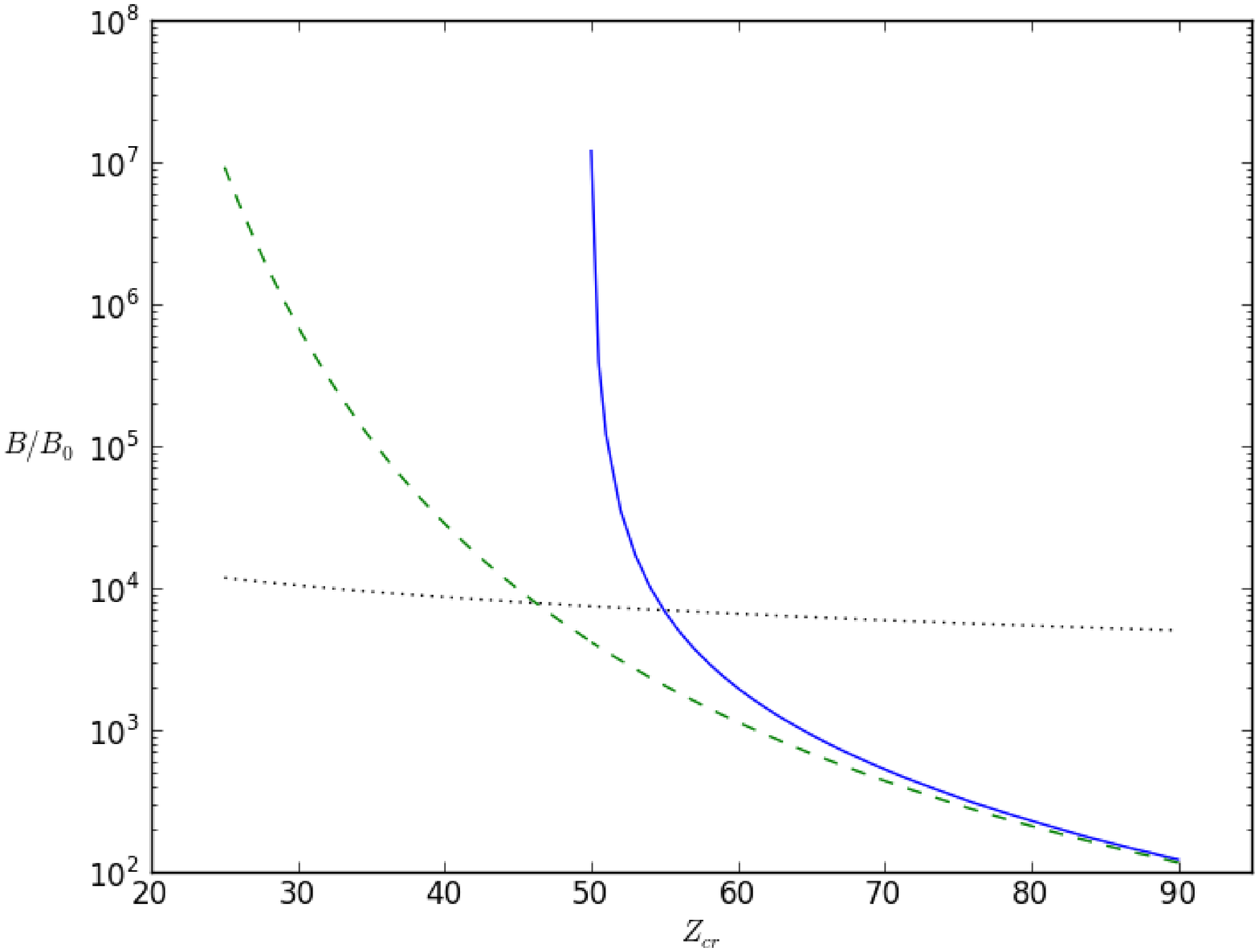}

Fig. 7. {\em The values of $B^{Z}_{cr}$: a) without screening according
    to (\ref{eq:ORS32}), dashed (green) line; b) numerical results
    with screening, solid (blue) line. The dotted (black) line corresponds
    to the field at which $a_{H}$ becomes smaller than the size of the
    nucleus.}

\end{center}

From Tables \ref{tab:1}, \ref{tab:3}, and \ref{table4} we see that (\ref{114}) is very good in
describing the dependence of the energy on the magnetic field; at
least a numerical integration produces almost identical results.
In Table \ref{table6} we demonstrate several cases where the accuracy of
(\ref{114}) is not that good. It happens at low $B/B_0$ since
the matching condition $B > 4m_e^2/(e(Ze^{2})^2)$ fails and when
$\varepsilon_{0}$ is relativistic. However, $B$ should not be too low
to make the adiabaticity condition $a_B \gg
a_H$, or $B \gg (Ze^2)^2 m_e^2/e$ applicable.
\begin{table}[h!]
\caption{\label{table6}Values of $\varepsilon_{0}/m_{e}$ at $B/B_0=5$.}
\begin{center}
\begin{tabular}{||c||c|c||}
\hline
$Z$   & Eq. (\ref{114})        & Numerical results   \\
${~}$ &       (Dirac)        &  (Dirac)            \\
\hline
90  &0.2050     &0.2512         \\
80  &0.3096     &0.3539         \\
70  &0.4139     &0.4542         \\
60  &0.5171     &0.5516         \\
50  &0.6185     &0.6454         \\
40  &0.7165     &0.7349         \\
30  &0.8086     &0.8188         \\
20  &0.8914     &0.8952         \\
10  &0.9596     &0.9601         \\
 1  &0.998745   &0.998745       \\
\hline
\end{tabular}
\end{center}
\end{table}

Textbooks \cite{greiner} contain detailed consideration of the phenomenon of critical charge.

\section{Conclusions}
An analytical formula for the Coulomb potential $\Phi(z)$ in a
superstrong magnetic field has been derived. It reproduces the results
of the numerical calculations made in \cite{SU} with good accuracy.
Using it, an algebraic formula for the energy spectrum of
the levels of a hydrogen atom
originating from the lowest Landau level in a superstrong $B$ has
been obtained. The
energies start to deviate from those obtained without taking
the screening of the Coulomb potential into account  at $B\ga
3\pi m_e^2/e^3 \approx 6 \cdot 10^{16}$ gauss and the energy of
ground state in the limit $B\longrightarrow\infty$ remains finite.

A magnetic field plays a double role in the critical charge
phenomenon. By squeezing the electron wave function and putting it in the
domain of a stronger Coulomb potential it diminishes
the value of the critical charge substantially \cite{ORS}. However, for
nuclei with $Z < 52$ to become critical such a strong $B$ is needed
that the screening of the Coulomb potential occurs and acts in the
opposite direction: the electron ground state energy freezes and the
nucleus remains subcritical in spite of growing $B$.

\vspace{0.5cm}
I am very grateful to Bruno Machet and Sergey Godunov for fruitful collaboration.

I am partially supported  by the RFBR grants
11-02-00441 and 12-02-00193, by the grant of the Russian Federation
government 11.G34.31.0047, and by the grant NSh-3172.2012.2.

\end{document}